\documentstyle[aps,prl,twocolumn,epsf]{revtex}
\begin{document}
\draft
\twocolumn[\hsize\textwidth\columnwidth\hsize
\csname @twocolumnfalse\endcsname
\title{Gaussian Tunneling Model of $c$-Axis Twist Josephson Junctions}
\author{A. Bille,$^1$, R. A. Klemm$^2$, and K.
Scharnberg,$^1$}
\address{$^1$I. Institut f{\"u}r Theoretische Physik, Universit{\"a}t
Hamburg, Jungiusstra{\ss}e 9, D-20355 Hamburg,
Germany}
\address{$^2$Materials Science Division, Argonne
National Laboratory, Argonne, IL
60439 USA}
\date{\today}
\maketitle
\begin{abstract}
We calculate the  critical current density $J^J_c$ for $c$-axis Josephson tunneling between identical high temperature superconductors   twisted an angle $\phi_0$ about the $c$-axis.   We model  the  tunneling matrix element squared as a Gaussian in the change of wavevector ${\bf q}$ parallel to the junction, $\langle|t({\bf q})|^2\rangle\propto\exp(-{\bf q}^2a^2/2\pi^2\sigma^2)$.  The $J^J_c(\phi_0)/J^J_c(0)$ obtained for the $s$- and extended-$s$-wave order parameters (OP's) are consistent with the  Bi$_2$Sr$_2$CaCu$_2$O$_{8+\delta}$ data of Li {\it et al.}, but only for  strongly incoherent tunneling, $\sigma^2\ge0.25$.  A $d_{x^2-y^2}$-wave OP is always inconsistent with the data.  In addition, we show that the apparent conventional sum rule violation observed by Basov {\it et al.} might be understandable in terms of incoherent $c$-axis tunneling, provided that the OP is not $d_{x^2-y^2}$-wave.  
\end{abstract}
\vskip0pt
\vskip0pt\vskip0pt
]
\narrowtext

Recently, there has been a resurgence of interest in the symmetry of the order parameter (OP) in the high temperature superconductors (HTSC). \cite{Kirtley,Mannhart,Tsuei,Giannetta,Gross,Li,MK,Kawayama} Although many  phase-sensitive experiments were interpreted as giving evidence for an OP in YBa$_2$Cu$_3$O$_{7-\delta}$ (YBCO) that was consistent with  the $d_{x^2-y^2}$-wave form,\cite{Kirtley,Mannhart}  the same type of  phase-sensitive experiments on the electron-doped HTSC Nd$_{1.85}$Ce$_{0.15}$CuO$_{4-y}$  and Pr$_{1.85}$Ce$_{0.15}$CuO$_{4-y}$  (PCCO) also were interpreted in terms of a $d_{x^2-y^2}$-wave OP, \cite{Tsuei} in apparent contradiction with other results,\cite{Giannetta,Gross} suggesting a possible problem with the former experiments. \cite{Kirtley,Mannhart}  In addition, new phase-sensitive experiments on Bi$_2$Sr$_2$CaCu$_2$O$_{8+\delta}$ (Bi2212) gave strong evidence that the OP contains an isotropic component, since it has a non-vanishing average over its Fermi surface. \cite{Li,MK,Kawayama,Yurgens}  First, $c$-axis Josephson tunneling between Bi2212 and either Pb or Nb demonstrated that an isotropic $s$-wave component of the OP exists in Bi2212. \cite{MK,Kawayama}  Second, $c$-axis tunneling across the junctions of Bi2212 intercalated with HgBr$_2$ has been studied using mesas. \cite{Yurgens}  In these experiments, increasing the $c$-axis spacing by 6.3{\AA} increased the normal state resistivity by a factor of 200.  In the superconducting state, $R_n$  and $I_c$ changed by comparable factors, but their product $I_cR_n\approx$ 10 mV, about half  the optimal value expected in the Ambegaokar-Baratoff (AB) model of purely incoherent $c$-axis tunneling between identical $s$-wave superconductors. \cite{Yurgens,AB}  Such behavior is very difficult to understand in terms of a $d_{x^2-y^2}$-wave OP. \cite{KARS}

More directly, a new phase-sensitive experiment which can test the symmetry of the OP over the entire range of temperatures $T$ below the superconducting transition temperature $T_c$ was performed. \cite{Li}  In this experiment, a single crystal of Bi2212 was cleaved in the $ab$-plane,  the two cleaves were twisted a chosen angle $\phi_0$ about the $c$-axis with respect to each other, and fused back together.  After lead attachment, the critical currents $I_c^J(T)$ and $I_c^S(T)$ across the twist junction and single crystals were measured, and their ratio was found to be  directly proportional to the ratio $A^J/A^S$ of their areas at $T/T_c=0.9$. \cite{Li}  Those authors then claimed  that (a) the intrinsic junctions and the twist junction behaved identically, (b) the $c$-axis tunneling is strongly incoherent, and (c) the OP contains an isotropic component, but not any purported $d_{x^2-y^2}$-wave component for $T<T_c$, except possibly below a second, unobserved phase transition.  \cite{Li} 

Since then, the group theoretic arguments upon which conclusion (c) were based have been published. \cite{KRS}  In addition, an exact calculation of the possible roles of coherent $c$-axis tunneling has been presented. \cite {AK} For the tight-binding Fermi surface generally thought to be applicable to Bi2212, \cite{Freeman,Olson}, it was shown that such coherent tunneling was inconsistent with the data, \cite{Li} even for an isotropic $s$-wave OP.   Since the claim (a) of Li {\it et al.} is just a statement of their experimental observations, and the claim (b) is clearly correct in the limit of purely incoherent tunneling, it remains to quantify precisely just how incoherent the tunneling must be in order to fit the data.  

Since Bi2212 behaves as a stack of weakly-coupled Josephson junctions, \cite{Mueller} the critical current density $J_c$ across each junction may be evaluated by neglecting the couplings between the other junctions. \cite{Li,KRS}  Specifically,  $J^J_c$ across the twist junction between layers 1 and 2 is given by
\begin{equation}
J_c^J=|4eT\sum_{\omega}\langle f^J({\bf k}-{\bf k}')F_1({\bf k}_J)F_2^{\dag}({\bf k_J}')\rangle|,
\end{equation}
where $\omega$ represents the Matsubara frequencies, $f^J({\bf k}-{\bf k}')$ is the spatial average of the tunneling matrix element squared, $\langle\ldots\rangle$ represents two-dimensional integrals over each of the  two first Brillouin zones (BZ's), and the wavevectors ${\bf k}_J$ and ${\bf k}_J'$ are obtained from ${\bf k}=(k_x,k_y)$ and ${\bf k}'=(k_x',k_y')$ by rotations of $\pm\phi_0/2$ about the $c$-axis, respectively.  The anomalous Green's functions
$F_n=\Delta_n/[\omega^2+\xi_n^2+|\Delta_n|^2]$ and $F_n^{\dag}=F_n^{*}$, where $\Delta_n({\bf k})$ and $\xi_n({\bf k})$ are the OP and quasiparticle dispersion on the $n$th layer, respectively.  For Bi2212, we assume the quasiparticle dispersions $\xi_n$ have identical tight-binding forms, differing only by the wavevector rotations,
\begin{eqnarray}
\xi_n({\bf k})&=&-t[\cos(k_xa)+\cos(k_ya)]\nonumber \\
 &&+t'\cos(k_xa)\cos(k_ya)-\mu,
\end{eqnarray}
where we take  $t=$306 meV, $t'/t=0.90$ and $\mu/t=-0.675$ to give a good fit to the Fermi surface of Bi2212, for which $\xi_n({\bf k}_F)=0$.  These values are  slightly different from those used previously. \cite{BRS}  A plot of this Fermi surface  is shown in Fig. 1.  We study  three OP's.  These are the isotropic $s$-wave, a constant, the $d_{x^2-y^2}$-wave, topologically equivalent to $\cos(k_xa)-\cos(k_ya)$, or  an extended-$s$-wave,  which we take to be the absolute magnitude of the particular $d_{x^2-y^2}$-wave form.  To obtain the particular $d_{x^2-y^2}$-wave form we used the repulsive interaction of the form \cite{Pines,Dahm}
\begin{equation}
V({\bf q})=-V_0\sum_{{\bf Q}=(\pm1,\pm1)\pi/a}\Gamma/[({\bf q}-{\bf Q})^2+\Gamma^2],\label{v}
\end{equation}
where $V_0=556$ meV and $\Gamma=0.1$.  The BCS equation is solved for $\Delta({\bf k})$ in terms of $V({\bf k-k}')$. The extended-$s$-wave OP was taken to have the absolute magnitude of the $d_{x^2-y^2}$-wave form obtained from Eq. (\ref{v}).   The isotropic $s$-wave OP was taken to have the maximum magnitude of it.  We included the umklapp terms which occur with rotated Fermi surfaces, but the effects of doing so were very small in all cases we studied.

\begin{figure}
%\vspace*{-1.5cm}
\epsfxsize=5.5cm
\centerline{\epsffile{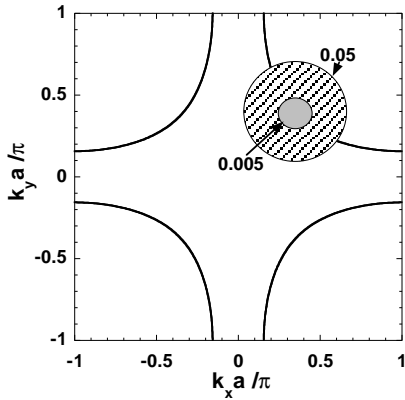}}
\vspace{0.3cm}
\caption{Schematic plot of the Bi2212 Fermi surface used in these calculations.  The concentric circles represent the portions of the BZ for which the tunneling strength is within $1/e$ of its maximum value, for dimensionless Gaussian parameters $\sigma^2$ of magnitude 0.005 and 0.05, as indicated.}\label{fig1}
\end{figure} 

Previously, we investigated the case in which $f^J({\bf k}-{\bf k}')\propto\delta({\bf k}-{\bf k}')$, appropriate for coherent tunneling. \cite{AK}  We found that it made very little difference whether or not one included umklapp terms in the integrals in Eq. (1).   However, we found that for Fermi surfaces similar to that pictured in Fig. 1, it was not possible to obtain a quantitative fit to the data. \cite{Li,AK}  We then argued that for extremely incoherent tunneling, with $f^J({\bf k}-{\bf k}')=f_0^J$, a constant,  the $s$-wave and extended-$s$-wave OP's could both fit the data equally well.  However, we did not discuss the intermediate case in any detail, and that is our purpose here.  We therefore take the tunneling matrix element squared to be a {\it Gaussian} in the momentum change parallel to the junction,
\begin{equation}
f^J({\bf k}-{\bf k}')=f_0^J\exp[-({\bf k}-{\bf k}')^2a^2/2\pi^2\sigma^2].\label{fJ}
\end{equation}
Since $\sigma=0$ and $\sigma\rightarrow\infty$ result in purely coherent and incoherent tunneling,  $\sigma$ is a dimensionless parameter quantifying the incoherence of the tunneling.  In Fig. 1, the  shaded concentric circles for $\sigma^2=0.005$ and 0.050 are pictured for the case in which the tunneling on one side of the junction occurs from the intersection of the Fermi surface and the line connecting the $(0,0)$ and $(\pi,\pi)$ points in the BZ.  In the shaded regions,  $|{\bf k}-{\bf k}'|\le\sigma\pi\sqrt{2}/a$, so that $1/e\le f^J({\bf k}-{\bf k}')/f_0^J\le1$. 

In the case of a circular Fermi surface cross-section, as might be expected for electron doped HTSC,  we could write ${\bf k}=k_F(\cos\phi_{\bf k},\sin\phi_{\bf k})$, etc. and Eq. (\ref{fJ}) can be rewritten as
\begin{equation}
f^J(\phi_{\bf k},\phi_{{\bf k}'})=f^J(\gamma)\exp[\gamma\cos(\phi_{\bf k}-\phi_{{\bf k}'})],\label{fJcircular}
\end{equation}
 where $\gamma=(ak_F/\pi\sigma)^2$ and $f^J(\gamma)=f_0^J\exp(-\gamma)$.  This form is  identical to that of Graf {\it et al.}, provided that  $f_0^J=\exp(\gamma)/[\tau_{\perp}I_0(\gamma)]$, \cite{Graf} where $I_n(z)$ is a Bessel function and $1/\tau_{\perp}$ is an effective interlayer tunneling rate.    The denominator contains the normalization factor $I_0(\gamma)$.  Then just below $T_c$, $\Delta(T)\rightarrow0$, and
\begin{equation}
J_{c\zeta}^J=C_0|\langle f^J(\phi_{\bf k},\phi_{{\bf k}'})\Delta_{1\zeta}(\phi_{{\bf k}_{+}})\Delta_{2\zeta}(\phi_{{\bf k}_{-}'})\rangle_{\phi_{\bf k},\phi_{{\bf k}'}}|,
\end{equation}
where $C_0=em^2/(4T_c)$, $m$ is the in-plane effective mass, $\zeta=s,d,e$ indexes the OP's, and $\phi_{{\bf k}_{+}}=\phi_{\bf k}+\phi_0/2$, $\phi_{{\bf k}_{-}'}=\phi_{{\bf k}'}-\phi_0/2$.
  These OP's are $\Delta_{ns}=\Delta_0$,  $\Delta_{nd}(\phi)=\Delta_0\cos(2\phi)$, and $\Delta_{ne}(\phi)=|\Delta_0\cos(2\phi)|$, respectively.  We find 
\begin{eqnarray}
J_{cs}^J&=&C_0\Delta_0^2/\tau_{\perp},\\
J_{cd}^J&=&C_0\Delta_0^2|\cos(2\phi_0)|f_d(\gamma)/\tau_{\perp}, \\ 
\noalign{\rm and}
J_{ce}^J&=&C_0\Delta_0^2f_e(\gamma,\phi_0)/\tau_{\perp}, \\ 
\noalign{\rm where}
f_d(\gamma)&=&{{I_2(\gamma)}\over{2I_0(\gamma)}} \\
\noalign{\rm and}
f_e(\gamma,\phi_0)&=&\sum_{n=0}^{\infty}{{4(2-\delta_{n,0})\cos(4n\phi_0)I_{4n}(\gamma)}\over{\pi^2(4n^2-1)^2I_0(\gamma)}}.\label{fofgamma}
\end{eqnarray}   

In this simple model,  $J_{ce}^J/J_{cs}^J=f_e(\gamma,\phi_0)$ at $T_c$, which is periodic in $\phi_0$ with period $\pi/2$, satisfying $f(\gamma,\pm\pi/2-\phi_0)=f(\gamma,\phi_0)$.  In the incoherent limit,   $f_e(0,\phi_0)=(2/\pi)^2$, a constant.  However, in the coherent limit $f_e(\infty,\phi_0)=[|\sin(2\phi_0)|+(\pi/2-2|\phi_0|)\cos(2\phi_0)]/\pi$ in the domain $|\phi_0|\le\pi/2$. $f_e(\infty,\phi_0)$ has maxima of $1/2$ at $\phi_0=n\pi/2$ and  minima of $1/\pi$ at $\phi_0=(2n+1)\pi/4$ for integer $n$.  The $\phi_0$ and $\gamma$ dependencies of $f_e(\gamma,\phi_0)$ are shown in Figs. 2 and 3.   

\begin{figure}
%\vspace*{-1.5cm}
\epsfxsize=5.5cm
\centerline{\epsffile{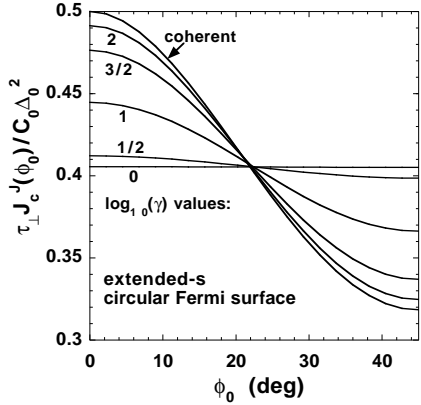}}
\vspace{0.3cm}
\caption{Plots of $f_e(\gamma,\phi_0)=\tau_{\perp}J_{ce}^J(\phi_0)/C_0\Delta_0^2$ just below $T_c$ for the extended-$s$-wave OP  $\Delta_0|\cos(2\phi_{\bf k})|$ on a circular Fermi surface, as a function of $\phi_0$ in degrees, at the $\gamma$ values $10^x$, where $x=0, 1/2, 1, 3/2, 2$, and for the coherent limit $\gamma\rightarrow\infty$.}\label{fig2}
\end{figure} 

However, for all $\gamma$ values,   $J_{cd}^J\propto|\cos(2\phi_0)|$, which vanishes  at $\phi_0=45^{\circ}$, as shown in Fig. 4.   In addition, $J_{cd}^J(\phi_0=0)$ differs from the $s$-wave case by the additional factor $f_d(\gamma)=I_2(\gamma)/[2I_0(\gamma)]$.  In the coherent limit, $\gamma\rightarrow\infty$, $f_d(\gamma)\rightarrow1/2$, but in the incoherent limit $\gamma\ll1$, $f_d(\gamma)\rightarrow\gamma^2/16$. 
 $f_e(\gamma,0)$ also decreases from $1/2$ at $\gamma\rightarrow\infty$ with decreasing $\gamma$, but  not nearly as dramatically, approaching the constant value $(2/\pi)^2$ as $\gamma\rightarrow0$.   The functions  $f_d(\gamma)$ and $f_e(\gamma,0)$  are compared in Fig. 3, where they are plotted as functions of $1/\gamma$ for clarity.  

\begin{figure}
%\vspace*{-1.5cm}
\epsfxsize=5cm
\centerline{\epsffile{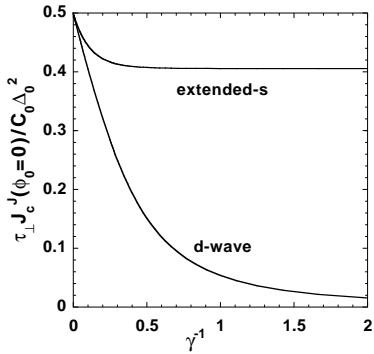}}
\vspace{0.3cm}
\caption{ Plots of $\tau_{\perp}J_c^J/C_0\Delta_0^2$ at $T_c$ and $\phi_0=0$, which is $f_e(\gamma,0)$ and $f_d(\gamma)$ for the extended-$s$-  and $d$-wave  OP's on the circular Fermi surface, respectively, versus  $1/\gamma$. }\label{fig3}
\end{figure}

We now consider the full calculations with the Fermi surface shown in Fig. 1, and Gaussian tunneling, Eq. ({\ref{fJ}).  We  performed explicit calculations at $T/T_c=0.5, 0.9$.   In Figs. 4-6, we  plotted our $J_c^J(\phi_0)/J_c^J(0)$ results at $T/T_c=0.5$  for the $d_{x^2-y^2}$-wave, extended-$s$-wave, and   isotropic $s$-wave OP's, respectively, for $0^{\circ}\le\phi_0\le45^{\circ}$.   Results for $T/T_c=0.9$ are very similar.  All results are periodic in $\phi_0$ with period $\pi/2$ and symmetric about $\phi_0=\pm\pi/4$.    Note that this presentation is equivalent to $J_c^J(\phi_0)/J_c^S$, since $J_c^S$ is the critical current density across the untwisted single crystal, which is itself a stack of essentially equivalent Josephson junctions.  Hence, $J_c^J(0)=J_c^S$.  In each case, we show the results for $\sigma^2=0, 0.005, 0.05, 0.25,$ and 0.50. 

\begin{figure}
%\vspace*{-1.5cm}
\epsfxsize=6cm
\centerline{\epsffile{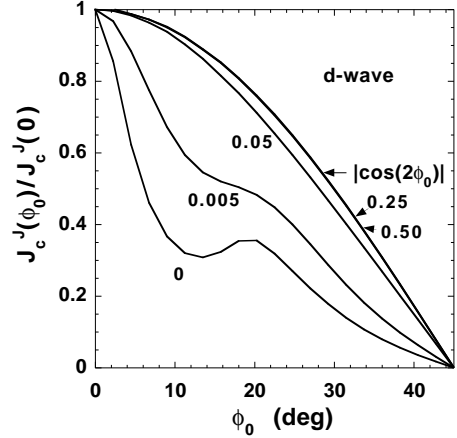}}
\vspace{0.3cm}
\caption{ Plots of $J_c^J(\phi_0)/J_c^J(0)$ at $T/T_c=0.5$ for twist junctions with the Fermi surface shown in Fig. 1, and a $d_{x^2-y^2}$-wave OP. The dimensionless Gaussian parameters $\sigma^2$ are indicated.  The curves for $\sigma^2=0.25$ and 0.50 are nearly  indistinguishable from the function $|\cos(2\phi_0)|$.}\label{fig4}
\end{figure}

\begin{figure}
%\vspace*{-1.5cm}
\epsfxsize=6cm
\centerline{\epsffile{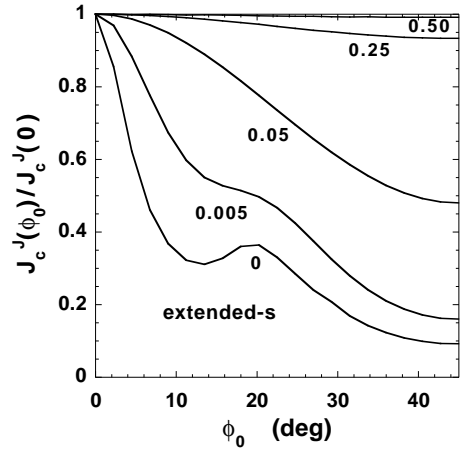}}
\vspace{0.3cm}
\caption{Plot of $J^J_c(\phi_0)/J^J_c(0)$ at $T/T_c=0.5$ for twist junctions with the Fermi surface shown in Fig. 1, and an extended-$s$-wave OP. The dimensionless Gaussian parameters $\sigma^2$ are indicated. }
\label{fig5}
\end{figure}

 For each OP, coherent tunneling   causes $J_c^J(\phi_0)/J_c^J(0)$ to decrease sharply as $\phi_0$ increases from $0^{\circ}$.  This is because the tight-binding Fermi surfaces are only identical for $\phi_0=0^{\circ}$, and the overlap of the Fermi surfaces changes dramatically from a continuous curve to a set of four points for $\phi_0>0^{\circ}$.  In addition, there are local maxima in $J_c^J(\phi_0)/J_c^J(0)$ at  about $\phi_0\approx 20^{\circ}$, which are absent for a circular Fermi surface cross-section. For this twist angle, the rotated Fermi surfaces begin to overlap at eight points.  This increased overlap is responsible for the more gradual decrease of $J_c^J(\phi_0)$ at larger $\phi_0$ values. We also note that the coherent tunneling curves for the extended-$s$- and $d$-wave OP's are nearly equal for $\phi_0\le30^{\circ}$.  This is also true for the $\sigma^2=0.005$ curves.  However, the $d$-wave curves are distinctly different from the $s$-wave and extended-$s$-wave curves in the vicinity of $\phi_0=45^{\circ}$, where the $d$-wave curves all vanish by symmetry.  In addition,  the $d$-wave curves for $\sigma^2=0.25$ and $\sigma^2=0.50$ are essentially identical and  quantitatively in agreement with the function $|\cos(2\phi_0)|$.  This is the same $\phi_0$ dependence that we obtained analytically for  all $\gamma$ in the case of a circular Fermi surface cross-section. 

\begin{figure}
%\vspace*{-1.5cm}
\epsfxsize=6cm
\centerline{\epsffile{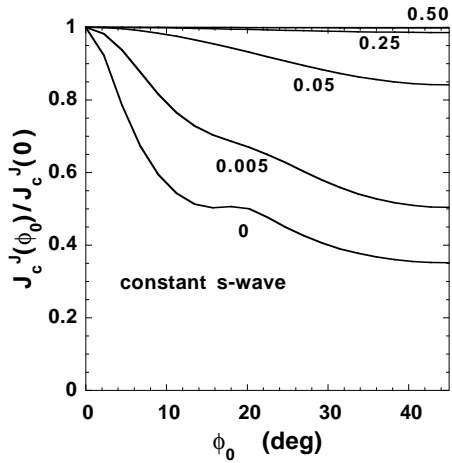}}
\vspace{0.3cm}
\caption{Plot of $J^J_c(\phi_0)/J^J_c(0)$ at $T/T_c=0.5$ for twist junctions with the Fermi surface shown in Fig. 1, and an isotropic $s$-wave order parameter.  The dimensionless Gaussian parameters $\sigma^2$ are indicated.}
\label{fig6}  \end{figure}

In comparing the cases of the isotropic and extended-$s$-wave OP's, we see that the main differences appear for rather coherent tunneling, for which the extended-$s$-wave curves have a more dramatic $\phi_0$-dependence.  However, both OP's are consistent with the data of Li {\it et al.} for $\sigma^2\ge0.25$, and neither one is consistent with the data for $\sigma^2\le0.05$.  By examining Fig. 1, we see that $\sigma^2=0.25$ corresponds to a tunneling strength satisfying $1/e\le f^J({\bf k}-{\bf k}')/f_0^J\le1$ over  39{\%} of the first BZ, which is strongly incoherent. Thus, these calculations imply that it is impossible to distinguish the isotropic-$s$-wave from the extended-$s$-wave OP's in the $c$-axis twist junction experiments.  We can only say that there must be an isotropic $s$-wave OP component for $T<T_c$, which is projected out from the OP anisotropy, if any, by  averaging over  a large portion of  the BZ.

Finally, we have studied the reductions in the product $I_cR_n$ of the critical current times the resistance across a junction from the AB limit ($I_cR_n(T)|_{AB}$) for the case of an extended-$s$-wave OP as $\sigma\rightarrow\infty$.   We first define
\begin{equation}
I_cR_n(T)=C(T/T_c)I_cR_n(T)|_{AB},
\end{equation}
where the AB limit curve corresponds to the isotropic $s$-wave case with $\sigma\rightarrow\infty$ for any Fermi surface.
 A few limits  can be investigated analytically.  For a circular Fermi surface cross-section,   we can analytically evaluate $I_cR_n$ for this model both at $T_c$ and as $T\rightarrow0$.  We find $C(1)=(2/\pi)^2\approx0.405$, as above, and $C(0)=(2/\pi)^2\ln 4\approx 0.562$.

Similar results can be obtained for   different Fermi surfaces.  For the Fermi surface shown in Fig. 1, we find numerically that $ C(0.9)=0.416$, $C(0.5)= 0.465$, and $C(0)=0.572$.  For the slightly different Fermi surface studied elsewhere, \cite{AK,BRS} with $t'/t=1.3$ and $\mu/t=-0.6$, we found $C(0.9)=  0.400$, $C(0.5)=0.450$, and $C(0)=0.578$.   Thus,  the result of Yurgens {\it et al.} that $I_cR_n(T=0) \approx$ 10 mV for the HgBr$_2$-intercalated Bi2212 is in rather good  agreement with that expected for an extended-$s$ OP. \cite{Yurgens}

We remark that there have been a number theories of the HTSC that relied upon coherent tunneling in the $c$-axis, with a ${\bf k}$-dependent matrix element, between each of the layers. \cite{Andersen,XW}   This model was shown previously to give the worst agreement with the experiment of Li {\it et al.} of all of the coherent tunneling models studied, even for an $s$-wave OP. \cite{Li,AK}  Here we showed that the experiment is easy to understand if the interlayer tunneling is strongly incoherent, provided that the OP integrates to a non-vanishing value on the two-dimensional Fermi ``surface''.  

Although there have not yet been many infrared reflectance measurements on Bi2212, the available experiments strongly suggest that the $c$-axis tunneling is not metallic.\cite{Tajima} In that work, not only was no Drude edge above $30$cm$^{-1}$ seen in the $c$-axis conduction in the normal state, none was seen in the superconducting state as well.  In addition, there is now strong evidence at all  underdoped HTSC have incoherent $c$-axis normal state tunneling, with the sole exception of  YBCO with $\delta\le0.15$.\cite{Tamasaku,Basov,Tajima2,Katz}  This is also the case in the recent measurements on the electron doped material NCCO. \cite{Basov2} Although above $T_c$, the non-metallic behavior is clearly seen, below $T_c$,  metallic-like behavior reminiscent of a Drude edge for wavevectors in the range 10 to 200 cm$^{-1}$ has been observed, \cite{Basov}   although none has yet been seen in Bi2212. \cite{Tajima} This metallic-like behavior  is most likely associated with the $c$-axis supercurrent. \cite{Basov}  Thus, if Bi2212 were to behave the same as the other materials with an incoherent $c$-axis normal state conduction, one would expect it to show this same metallic-like behavior below $T_c$.  

There have been two schools of thought on this issue.  One is that the interlayer tunneling processes change dramatically from incoherent to coherent below $T_c$.  \cite{Bulaevskii} Our analysis of the $c$-axis twist experiments of Li {\it et al.}  strongly contradicts this idea, since the quasiparticle tunneling below $T_c$ must be strongly incoherent. The second school holds that there is no quasiparticle tunneling below $T_c$ due to an ``orthogonality catastrophe'', but that the only tunneling process occurs by the simultaneous tunneling of pairs. \cite{PWA}  However, the fact that HgB$_2$-intercalated Bi2212 has nearly the same $T_c$ as does unintercalated Bi2212 argues strongly that this interlayer pair tunneling model for the occurrence of superconductivity is not correct. \cite{Yurgens}

Thus, if the superconductivity arises from intralayer pairing, then the tunneling process must be tunneling by individual quasiparticles.  The $c$-axis twist experiments provide strong evidence that this tunneling is incoherent, meaning that the momenta of the quasiparticles parallel to the layers before and after the tunneling process are not correlated, or random.  This is in addition to any possible changes in the kinetic energy that might also occur. \cite{Basov}

With incoherent normal state $c$-axis tunneling, it is extremely difficult to obtain a significant amount of $c$-axis critical current with an OP that averages to zero on the Fermi ``surface''.  Since there is a very strong consensus that the superconducting critical current in all HTSC is three-dimensional, incoherent tunneling in the $c$-axis direction necessarily implies that the OP must have an $s$-wave component, so that its Fermi ``surface'' average is non-vanishing at all $T\le T_c$.  This is also true in the organic layered superconductor $\kappa$-(ET)$_2$Cu[N(CN)$_2$]Br, which has  recently been shown to have extremely incoherent $c$-axis propagation, both above and below $T_c$. \cite{Timusk}
Combined with this supporting evidence that the $c$-axis tunneling in most HTSC is predominantly incoherent, the twist experiments of Li {\it et al.}    provide very strong  evidence of an $s$-wave superconducting OP in the HTSC.  

Very recently, a simple model for the apparent violation of the conventional sum rule observed by Basov 
{\it et al.} was investigated theoretically by Kim and Carbotte (KC).\cite{Carbotte}  Those workers 
assumed a $d_{x^2-y^2}$-wave OP, and an incoherent interlayer tunneling matrix element squared of the 
form $f^J(\phi_{\bf k},\phi_{{\bf k}'})=|V_0|+|V_1|^2\cos(2\phi_{\bf k})\cos(2\phi_{{\bf k}'})$. By 
equating $1/\lambda_c^2$ derived from the conductivity and the superfluid density $\rho_s$,  each 
evaluated to lowest order in $f^J$, they wrote,\cite{Carbotte}
\begin{equation}
\delta N_{\zeta}(T/T_c)={1\over2}+{{\sum_{\omega}\langle f^J(\phi_{\bf k},\phi_{{\bf k}'})[1-
\omega_{\zeta\bf k}\omega_{\zeta{\bf k}'}]\rangle}\over{2\sum_{\omega}
\langle f^J(\phi_{\bf k},\phi_{{\bf k}'})\delta_{\zeta\bf k}\delta_{\zeta{\bf k}'}\rangle}},\label{deltaN}
\end{equation}
where $\delta N_{\zeta}=(N_n-N_{s\zeta})/\rho_{s\zeta}$, $\omega_{\zeta\bf k}=\omega/\Omega_{\zeta\bf k}$, $\omega_{\zeta{\bf 
k}'}=\omega/\Omega_{\zeta{\bf k}'}$, $\delta_{\zeta\bf k}=\Delta_{\zeta}(\phi_{\bf k})/\Omega_{\zeta\bf 
k}$, $\delta_{\zeta{\bf k}'}=\Delta_{\zeta}(\phi_{{\bf k}'})/\Omega_{\zeta{\bf k}'}$, $\Omega_{\zeta\bf 
k}=[\omega^2+\Delta^2_{\zeta}(\phi_{\bf k})]^{1/2}$, and $\Omega_{\zeta{\bf 
k}'}=[\omega^2+\Delta^2_{\zeta}(\phi_{{\bf k}'})]^{1/2}$.  It is easy to show that for coherent tunneling $\delta N_{\zeta}(T/T_c)=1$ for any OP, for all $T/T_c\le1$.  It is also easy to show that $\delta N_s(T/T_c)=1$, regardless of the form of $f^J(\phi_{\bf k},\phi_{{\bf k}'})$. For the $d$-wave case,  KC showed that incoherent tunneling gives a 
conductivity sum rule violation that is strong and of the wrong sign.  \cite{Basov,Carbotte,Ioffe}  
They gave a lower limit on the violation, based upon restrictions of the parameters $V_0$ and $V_1$.   However, KC did not calculate the extended-
$s$-wave case, which is actually very interesting to do. 

We studied the one-parameter model  of Graf 
{\it et al.}, \cite{Graf}, which interpolates smoothly between coherent and incoherent tunneling. In this model, we can analytically perform the calculations at $T_c$ for the $\zeta=d,e$ cases.  
We find that $\delta 
N_d(1)=[2+1/f_d(\gamma)]/4$ and $\delta N_e(1)=[2+1/f_e(\gamma,0)]/4$. 
In the coherent limit $\gamma\rightarrow\infty$, all three OP's give no sum rule violation, as expected.  However, in the 
incoherent limit $\gamma\rightarrow0$, $\delta N_d(1)\rightarrow4/\gamma^2$, which diverges strongly, 
and $\delta N_e(1)\rightarrow(1+\pi^2/8)/2\approx 1.117$. As $T\rightarrow0$ in the coherent limit 
$\gamma\rightarrow\infty$, we again have $\delta N_{\zeta}(0)=1$ for $\zeta=s,d,e$, as expected.  As $T\rightarrow0$ 
in the incoherent limit $\gamma\rightarrow0$, for both 
$\zeta=d,e$, we can evaluate the denominator in Eq. (\ref{deltaN}) exactly, and the numerator numerically.    
We find $\delta N_e(0)\rightarrow 1.087$ and $\delta N_d(0)\rightarrow\infty$. We thus conclude that 
none of the three OP's gives a $\delta N<1$, but the $d$-wave case is by far the worst, especially in the limit 
of strongly incoherent tunneling.  Thus, it is  likely to be much easier to construct a theory that can 
incorporate both incoherent interlayer tunneling and a $\delta N<1$ if the OP is an anisotropic $s$-wave 
one, and not $d$-wave.   In particular, theories based upon two gaps, a non-superconducting pseudogap and 
a superconducting  gap, appear likely to satisfy the experimental sum rule violation, provided that the OP 
describing the superconducting state is either $s$-wave or extended-$s$-wave.

In conclusion, we found that the $c$-axis twist experiments of Li {\it et al.} provide compelling evidence that the $c$-axis tunneling in Bi2212 is strongly incoherent. \cite{Li} As a consequence, the experiment cannot distinguish between  an isotropic $s$-wave  and an extended-$s$-wave OP.  However, the purported $d_{x^2-y^2}$-wave OP can be ruled out in Bi2212.   The other HTSC and organic layered superconductors which also have such incoherent $c$-axis tunneling also cannot have a $d_{x^2-y^2}$-wave OP.  Such incoherent $c$-axis tunneling is likely to be consistent with the available $c$-axis infrared reflectivity measurements, provided that the OP is not of the $d_{x^2-y^2}$-wave form.

%\acknowledgments

The authors  thank G. Arnold, R. Kleiner, Qiang Li, and M. M{\"o\ss}le for useful discussions.
This work was supported by 
USDOE-BES  Contract No. W-31-109-ENG-38,
by NATO  Collaborative Research Grant No. 960102,
and by the DFG through
the Graduiertenkolleg ``Physik nanostrukturierter
Festk\"orper.''

\end{document}